\documentclass[twocolumn,showpacs,pre,floatfix,superscriptaddress]{revtex4}
\usepackage{color} 
\usepackage{tabularx} 
\usepackage{epsfig}
\usepackage{amsmath} 
\usepackage{amssymb} 
\usepackage{graphicx}
\usepackage{wasysym}
\usepackage{times}
\usepackage{float}

\begin{document}

\title{Fair sampling of ground-state configurations of binary optimization
problems}

\author{Zheng Zhu}
\affiliation {Department of Physics and Astronomy, Texas A\&M University,
College Station, Texas 77843-4242, USA}

\author{Andrew J.~Ochoa}
\affiliation {Department of Physics and Astronomy, Texas A\&M University,
College Station, Texas 77843-4242, USA}

\author{Helmut G.~Katzgraber}
\affiliation{Microsoft Quantum, Microsoft, Redmond, Washington 98052, USA}
\affiliation {Department of Physics and Astronomy, Texas A\&M University,
College Station, Texas 77843-4242, USA}
\affiliation{Santa Fe Institute, 1399 Hyde Park Road, Santa Fe, New Mexico 
87501, USA}

\date{\today}

\begin{abstract}

Although many efficient heuristics have been developed to solve binary
optimization problems, these typically produce correlated solutions for
degenerate problems. Most notably, transverse-field quantum
annealing---the heuristic employed in current commercially available
quantum annealing machines---has been shown to often be exponentially
biased when sampling the solution space. Here we present an approach to
sample ground-state (or low-energy) configurations for binary
optimization problems. The method samples degenerate states with almost
equal probability and is based on a combination of parallel tempering
Monte Carlo with isoenergetic cluster moves. We illustrate the
approach using two-dimensional Ising spin glasses, as well as spin
glasses on the D-Wave Systems quantum annealer chimera topology.
In addition, a simple heuristic to approximate the number of solutions
of a degenerate problem is introduced.

\end{abstract}

\pacs{75.50.Lk, 75.40.Mg, 05.50.+q, 64.60.-i}

\maketitle

\section{Introduction}

Quantum annealing
\cite{finnila:94,kadowaki:98,brooke:99,farhi:01,santoro:02,das:05,santoro:06,das:08,morita:08}
and, in particular, quantum annealing machines have ignited an
ever-increasing interest in algorithms used in statistical physics to
solve hard combinatorial industrial optimization problems, as well as
related applications. While there has been an extensive body of work
attempting to discern if the D-Wave Systems special-purpose quantum
annealing machine can outperform algorithms on conventional CMOS
hardware
\cite{dickson:13,pudenz:13,smith:13,boixo:13a,albash:15a,ronnow:14a,katzgraber:14,lanting:14,santra:14,shin:14,boixo:14,albash:15,albash:15a,katzgraber:15,martin-mayor:15a,pudenz:15,hen:15a,venturelli:15a,vinci:15,zhu:16,mandra:16b,mandra:17a,mandra:18},
there have been only a few studies
\cite{boixo:13a,albash:14,king:16,mandra:17,koenz:18y} attempting to
characterize the sampling ability of quantum annealing. Initial studies
\cite{matsuda:09,mandra:17} suggested that transverse-field quantum
annealing with stoquastic drivers result in biased solution
distributions for degenerate problems. However, more recently, it was
shown \cite{koenz:18y} that even with high-order drivers the sampling
bias can be removed only in special cases.

Many industrial applications rely more on a broad solution pool then on
the minimum of the cost function, with some prominent examples being
propositional model counting and related problems
\cite{jerrum:86,gomes:08,gopalan:11}, SAT-based probabilistic membership
filters \cite{weaver:14,schaefer:78,douglass:15,herr:17}, machine
learning applications \cite{hinton:02,eslami:14}, or simply estimating
the ground-state entropy of a degenerate system. In addition, having
multiple solutions to a given problem might allow for the inclusion of
constraints in a post-processing step.  Here we demonstrate that Monte
Carlo methods paired with cluster updates can result in algorithms that
asymptotically sample ground-states fairly.

Classical Monte Carlo heuristics based on thermal annealing are known to
almost uniformly sample all ground-state and low-lying excited state
configurations \cite{moreno:03,wang:15}. Studies of three-dimensional
diluted Ising antiferromagnets in a field and three-dimensional Ising
spin glasses show that parallel tempering Monte Carlo
\cite{hukushima:96} is more efficient than simulated annealing
\cite{kirkpatrick:83} at finding spin-glass ground-state configurations
with near-equal probability \cite{moreno:03,comment:pa}. 
Isoenergetic cluster  moves (ICM) \cite{zhu:15}, related to Houdayer's
cluster updates \cite{houdayer:01}, introduced for Ising spin glasses
significantly speed up thermalization on quasi-two-dimensional
topologies, such as D-Wave's Chimera graph.
The combination of low-temperature parallel tempering (PT)
Monte Carlo and the rejection-free isoenergetic cluster moves (PT+ICM)
allow for a wide-spread sampling of search space and help escape local
minima separated by large energy barriers. Here we demonstrate that
isoenergetic cluster moves paired with parallel tempering Monte Carlo
(PT+ICM) enhance the fair sampling of ground-state configurations
for spin-glass problems better than the previous PT gold standard. We
illustrate the approach using two-dimensional Ising spin glasses on a
square lattice, as well as the Chimera graph. Higher-dimensional
problems can be embedded in lower-dimensional graphs where PT+ICM is
more efficient via, e.g., minor embedding \cite{choi:08,choi:11}.

The paper is organized as follows. In Sec.~\ref{sec:model} we introduce
a quality metric for fair sampling, as well as a detailed
description of a fair-sampling algorithm using ICM. Following that, we
present numerical results in Sec.~\ref{sec:numerics} for both PT, as
well as PT+ICM, and introduce an algorithm to approximate the number of
degenerate states for highly-degenerate problems. We conclude with a
discussion of our results.

\section{Model and Algorithm}
\label{sec:model}

To illustrate the improved sampling of PT+ICM over PT, we start with an
Ising spin-glass model on a nonplanar Chimera graph \cite{bunyk:14}.
Its nonplanar topology makes finding ground states of random Ising spin
glasses worst-case NP-hard. The Hamiltonian for the spin-glass model is
given by
\begin{equation}
{\mathcal H} = -\sum_{i<j}^N J_{ij} s_i \,s_j, 
\label{:ham}
\end{equation}
where $s_i \in\{\pm 1\}$ are Ising spins and the couplers $J_{ij}$ are
drawn for this study from three discrete distributions: $\{\pm1,\pm2,\pm4\}$,
$\{\pm5,\pm6,\pm7\}$ and $\{\pm1\}$).  The couplers are selected based
on the range of ground-state degeneracy we can handle with our
high-performance computing cluster; i.e., the less symmetries between
the  different coupler values, the smaller the ground-state degeneracy.

\subsection{Assessing optimal sampling}

Suppose $n$ is the total number of times that ground states are found
for an instance with ground-state degeneracy $G$. The probability
distribution for finding any particular ground-state configuration
follows a \textit{binomial} distribution. For theoretically perfect
sampling, if $p=1/G$ is the probability of finding a state and $q=1-p$
is the probability of failure in a given trial, then the expected number
of successes in $n$ trials is $e=np$ and the variance of the binomial
distribution is $\sigma^2 = npq$. Therefore, the theoretical relative
standard deviation given by sampling a finite set of random uncorrelated
numbers $Q_{\rm th}$ is given by
\begin{equation}
Q_{\rm th}=\sigma/e=\sqrt{(1-p)/np}=\sqrt{(G-1)/n}.
\label{:rv}
\end{equation}
Assuming that the states are uncorrelated (which is a safe assumption
for large $G$), an algorithm is said to be optimal (sampling fairly) if
the numerical relative standard deviation of the frequency of
ground-state configurations $Q_{\rm num}$ determined experimentally is
close or equal to the theoretical value $\sqrt{(G-1)/n}$ (or $Q_{\rm
num}/Q_{\rm th}=1$). In practice, $Q_{\rm num}$ for any algorithm is
almost always greater than the theoretical value $Q_{\rm th}$, due to a
limited number of measurements via e.g., limited computing resources.

\subsection{PT+ICM for fair sampling}

Our implementation of PT+ICM for sampling purposes can be summarized as
follows:
\begin{enumerate}

\item{Run $N_T$ replicas of the system at a range of temperatures
\{$T_1,T_2,...,T_{N_T}$\}, with each set consisting of $M=4$ copies of
the system at the same temperature, thus $4 \times N_T$ copies of the
system with the same disorder are randomly initialized.}

\item{$N_{\rm sw}$ iterations are performed, each iteration consisting
of one Monte Carlo sweep, a parallel tempering update, and an
isoenergetic cluster move (for the lowest $N_{\rm hc}$ temperatures).}

\item{For the first $N_{\rm sw}/2$ iterations, keep track of the
lowest energies for the four replicas at the lowest temperatures.}

\item{After $N_{\rm sw}/2$ iterations, the lowest energies $E_1$, $E_2$,
$E_3$, and $E_4$ for the four replicas with the lowest temperatures are
compared, and if $E_1=E_2=E_3=E_4$, the ground-state energy has been
found with high confidence. Once this is the case, configurations at
this energy are recorded, as well as their frequency for the remaining
$N_{\rm sw}$/2 updates.}

\end{enumerate}
There is no guarantee that any solution obtained by this heuristic
method is the true optimum, or that we have found all configurations
that minimize the Hamiltonian. However, we choose to make sure each
configuration achieves a minimum number of $50$ hits in order to increase
our confidence that all accessible ground states have been found. The
simulation parameters are shown in Table \ref{tab:simparams_sampling}.

\begin{table}
\caption{
Parameters of the simulation: For each instance class and system size
$N$, we compute $N_{\rm sa}$ instances. $N_{\rm sw} = 2^b$ is the total
number of Monte Carlo sweeps for each of the $4 N_T$ replicas for a
single instance, $T_{\rm min}$ [$T_{\rm max}$] is the lowest [highest]
temperature simulated, and $N_T$ and $N_{\rm hc}$ are the number of
temperatures used in the parallel tempering method and in the
isoenergetic cluster algorithm, respectively.
\label{tab:simparams_sampling}
}
\begin{tabular*}{\columnwidth}{@{\extracolsep{\fill}} l c l l r l l l r }
\hline
\hline
Topology & Couplers & $N$ & $N_{\rm sa}$ & $b$ & $T_{\rm min}$ & $T_{\rm max}$ & $N_{T}$ &$N_{hc}$\\
\hline
2D & $\{\pm1,\pm2,\pm4\}$ & $144$ & $360$ & $24$ & $0.05$ & $3.05$ & $35$ &$35$\\
2D & $\{\pm1,\pm2,\pm4\}$ & $256$ & $360$ & $24$ & $0.05$ & $3.05$ & $35$ &$35$\\
2D & $\{\pm1,\pm2,\pm4\}$ & $576$ & $322$ & $24$ & $0.05$ & $3.05$ & $35$ &$35$\\
2D & $\{\pm1,\pm2,\pm4\}$ & $784$ & $232$ & $24$ & $0.05$ & $3.05$ & $35$ &$35$\\
2D & $\{\pm1,\pm2,\pm4\}$ & $1024$ & $370$ & $24$ & $0.05$ & $3.05$ & $35$ &$35$\\
\hline
Chimera  & $\{\pm1,\pm2,\pm4\}$ & $128$ & $360$ & $24$ & $0.05$ & $3.05$ & $35$ &$20$\\
Chimera  & $\{\pm1,\pm2,\pm4\}$ & $288$ & $360$ & $24$ & $0.05$ & $3.05$ & $35$ &$20$\\
Chimera  & $\{\pm1,\pm2,\pm4\}$ & $512$ & $360$ & $24$ & $0.05$ & $3.05$ & $35$ &$20$\\
Chimera  & $\{\pm1,\pm2,\pm4\}$ & $800$ & $360$ & $24$ & $0.05$ & $3.05$ & $35$ &$20$\\
Chimera  & $\{\pm5,\pm6,\pm7\}$ & $800$ & $976$ & $24$ & $0.10$ & $1.55$ & $30$ &$23$\\
Chimera  & $\{\pm1,\pm2,\pm4\}$ & $1152$ & $223$ & $24$ & $0.05$ & $3.05$ & $35$ &$20$\\
\hline
\end{tabular*}
\end{table}

\section{Numerical results}
\label{sec:numerics}

\begin{figure}[!htb]
\centering
\includegraphics[width=\columnwidth]{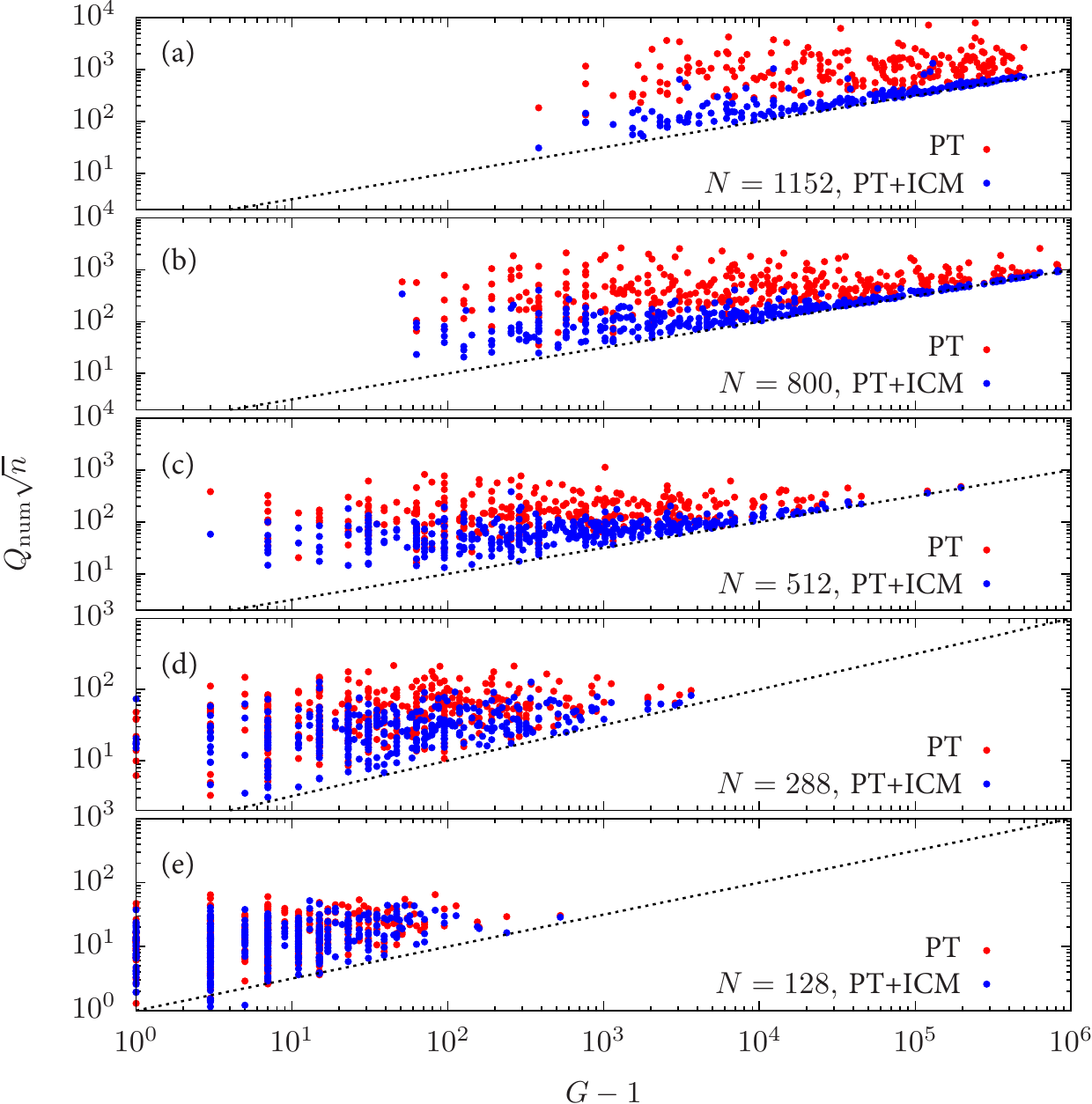}
\caption{
Scatter plot of $Q_{\rm num}\sqrt n$ as a function of the
ground-state degeneracy $G-1$ for different spin-glass instances with
different system sizes $N$ on a Chimera graph. The data points for PT+ICM
(blue/darker dots) are closer to the theoretical limit than those from PT
(red/lighter dots), and the improvement improves as the system size
increases. The dotted line represents ideal uniform sampling of
ground-state configurations, i.e., $Q_{\rm num}/Q_{\rm th}=1$.
Note that any other heuristic, such as simulated or quantum annealing 
would perform worse than PT \cite{moreno:03,mandra:17}. 
Data for (a) $N = 1152$,
(b) $N = 800$,
(c) $N = 512$,
(d) $N = 288$, and
(e) $N = 128$.
}
\label{fig:chimera_scatter}
\end{figure}

\begin{figure}[!htb]
\centering
\includegraphics[width=\columnwidth]{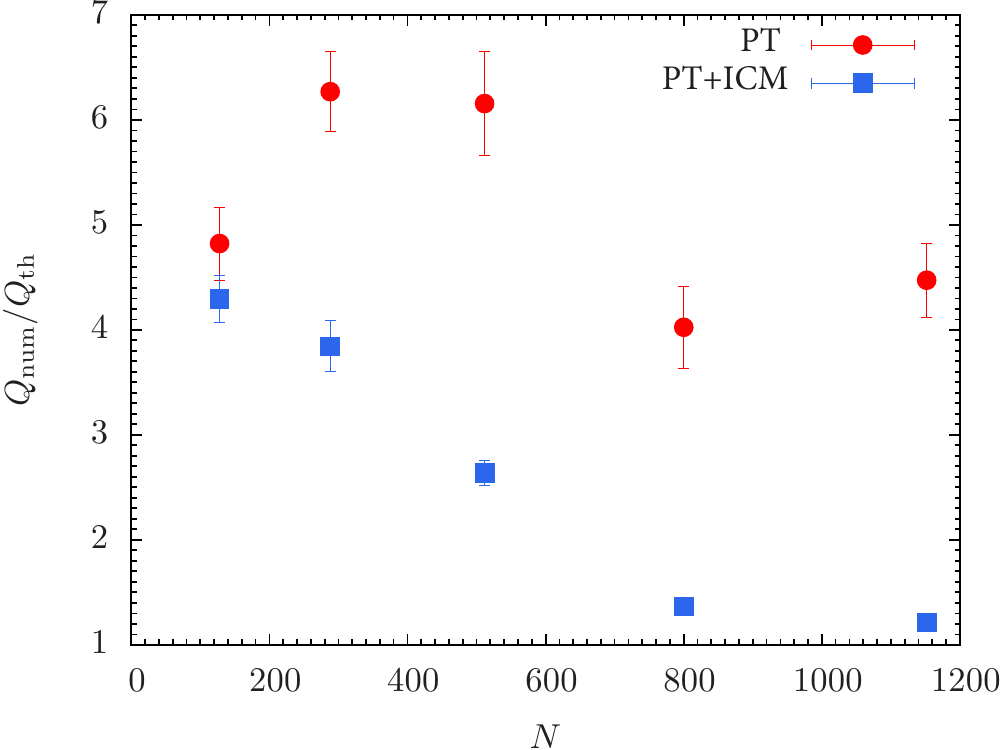}
\caption{
Median ratio $Q_{\rm num}/Q_{\rm th}$ for spin-glass instances on
Chimera as a function of the system size $N$. The data points show that
PT+ICM (blue squares) performs better than PT (red circles) for all
system sizes and the gain is more significant with increasing system
size. Statistical error bars are determined by a bootstrap
analysis.
}
\label{fig:chimera_ratio}
\end{figure}

To test whether PT+ICM can sample ground-state configurations with
near-equal probabilities, we multiply the numerical relative standard
deviation $Q_{\rm num}$ by $\sqrt n$ and plot $Q_{\rm num}\sqrt n$ as a
function of the ground-state degeneracy $G-1$.  Note that $Q_{\rm
th}\sqrt n$ is the square root of the ground-state degeneracy $G-1$, and
therefore the function $Q_{\rm th}\sqrt n=\sqrt{G-1}$ is a straight line
in logarithmic scale for both the horizontal axis ($G-1$) and the
vertical axis ($Q_{\rm th}\sqrt n$).

Figure \ref{fig:chimera_scatter} shows $Q_{\rm num}\sqrt n$ and $Q_{\rm
th}\sqrt n$ as a function of the ground-state degeneracy $G-1$ for
different spin-glass instances on a Chimera graph. As mentioned in the
previous paragraph, the quantity $Q_{\rm num}\sqrt n$ is almost always
greater than $Q_{\rm th}\sqrt n$ due to limited computational resources
\cite{comment:ratio_one}. However, an algorithm samples optimally if the
data from the numerical relative standard deviation are close to the
theoretical line. It is clear that the data for PT+ICM (blue/darker
color) are closer to a straight line than the data for PT (red/lighter
color), and the discrepancy between PT+ICM and PT seems to become
greater as the system size increases.

In Fig.~\ref{fig:chimera_ratio} we plot the median ratio $Q_{\rm
num}/Q_{\rm th}$ as a function of the system size $N$ for spin-glass
problems on a Chimera lattice. We emphasize that when the ratio becomes
unity an algorithm samples optimally.  The data show that PT+ICM (blue
squares) performs better than PT (red circles) and that the improvement
is more significant with increasing system size. In this work the
temperature set for the simulation is specifically optimized for
$N=1152$. Large median ratios $Q_{\rm num}/Q_{\rm th}$ for smaller
system sizes are due to the choice of temperature set. The statistical
error bars are determined by a bootstrap analysis using the following
procedure: For each system size $N$ and $N_{\rm sa}$ disorder
realizations, a randomly selected bootstrap sample of the $N_{\rm sa}$
disorder realizations is generated. The median ratio $Q_{\rm num}/Q_{\rm
th}$ is computed with this random sample. We repeat this procedure
$N_{\rm boot}=1000$ times for each system size to obtain an average and
error bar using these $N_{\rm boot}=1000$ data points.

\begin{figure}[!htb]
\centering
\includegraphics[width=\columnwidth]{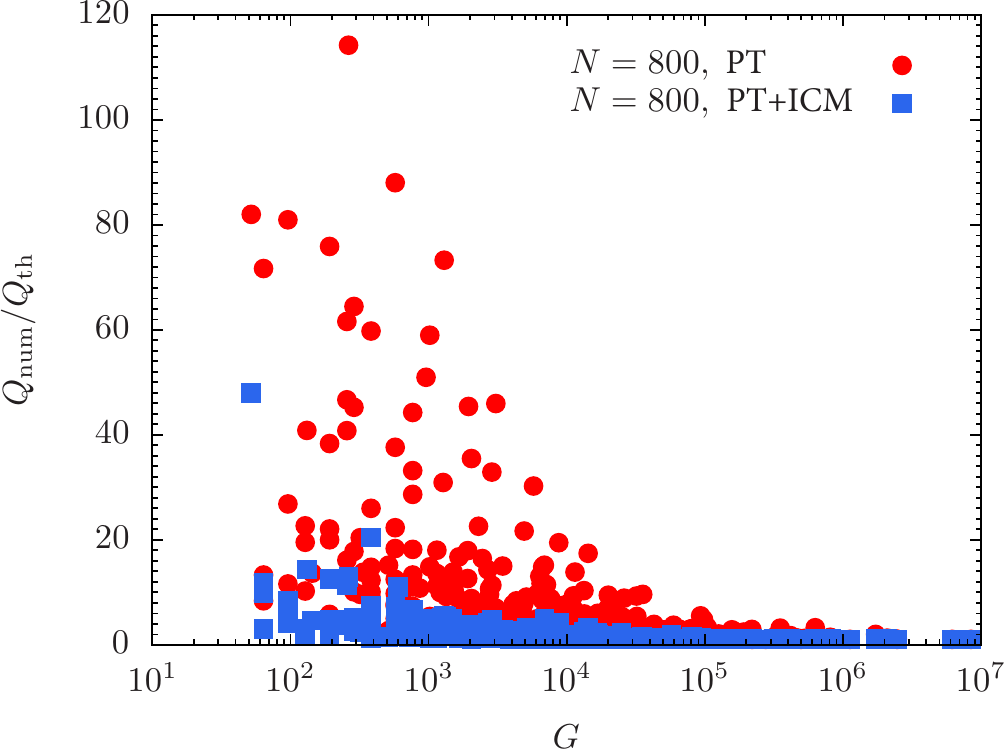}
\caption{
Scatter plot of $Q_{\rm num}/Q_{\rm th}$ as a function of the estimated
ground-state degeneracy $G$ for different spin-glass instances with
system size $N=800$ on a Chimera graph. Both data for PT and PT+ICM
suggest that the more ground-state configurations, the easier to sample
all ground-state configurations with near-equal probabilities using
these heuristics.
}
\label{fig:chimera_Q_degeneracy}
\end{figure}

\begin{figure}[!htb]
\centering
\includegraphics[width=\columnwidth]{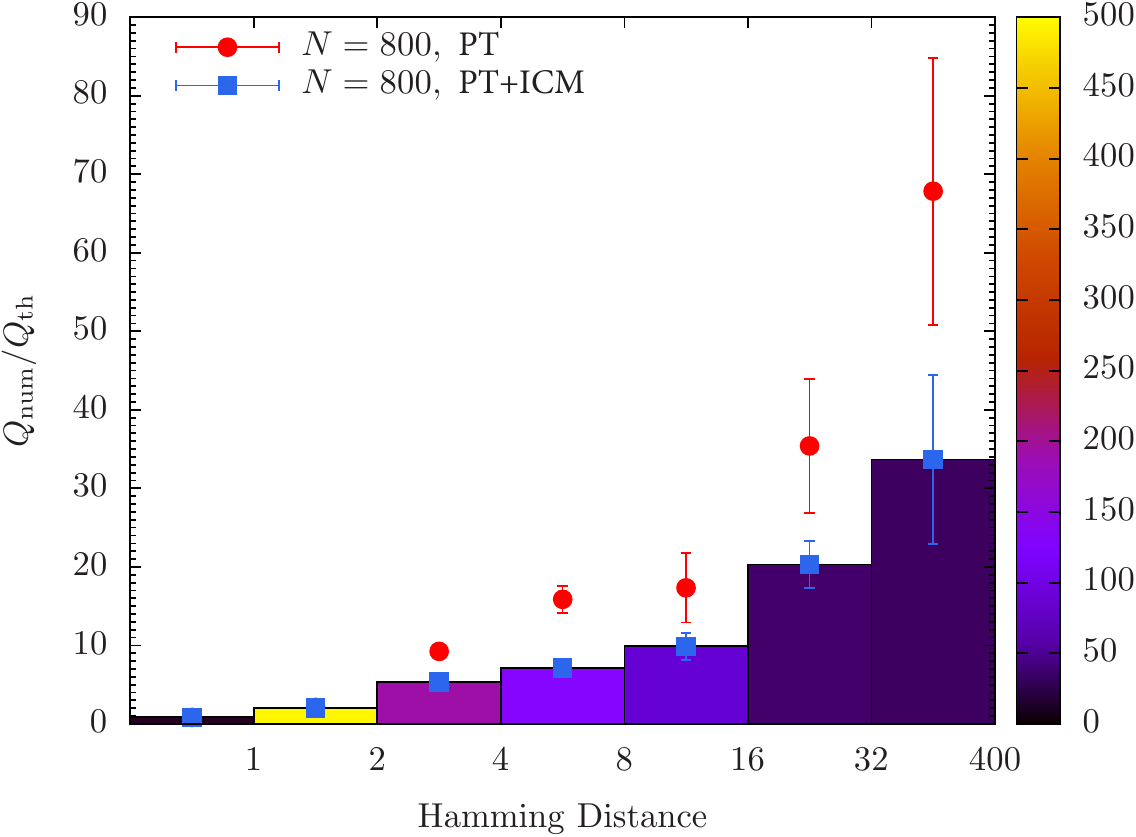}
\caption{
Median ratio $Q_{\rm num}/Q_{\rm th}$ as a function of Hamming distance
for different spin-glass Sidon instances ($J_{ij} \in
\{\pm5,\pm6,\pm7\}$) \cite{katzgraber:15} with system size $N=800$ and
degeneracy $G=2$ (up to spin reversal symmetry) on a Chimera graph.
Data for PT and PT+ICM suggest that the larger the Hamming distance
between ground-state configurations, the harder it is to sample all
ground-state configurations with near-equal probability and the more
PT+ICM improves fair sampling over PT. Note that the bar chart
represents median ratios $Q_{\rm num}/Q_{\rm th}$ between Hamming
distance $1-2$, $2-4$, $4-8$, $8-16$, $16-32$, and $32-400$,
respectively. The statistical error bars are determined by a bootstrap
analysis. Bars are color coded with the number of instances that have a 
particular Hamming distance. 
}
\label{fig:chimera_Q_hamming}
\end{figure}

\begin{figure}[!htb]
\centering
\includegraphics[width=\columnwidth]{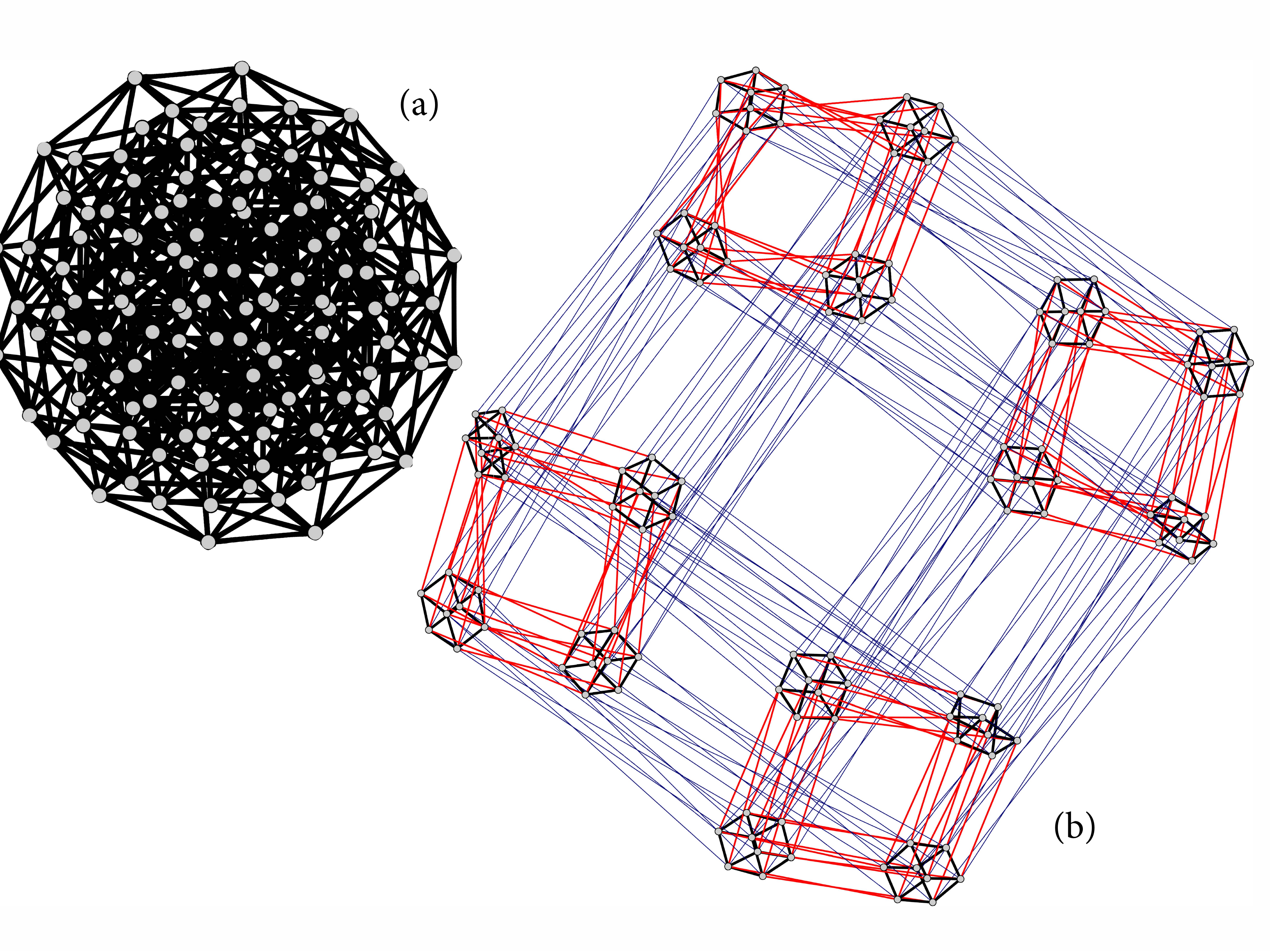}
\caption{
Two examples of ground-state configurations with different Hamming
distances on a Chimera graph for system size $N=128$. The lines
represents the distance between two binary strings (ground-state
configurations). Each dot in the figure represents a ground-state
configuration, black (thick) lines are $1$-bit differences, red lines 
(medium shade) are
$2$-bit differences, and lighter colors (light gray or blue) indicate an 
even greater difference. In the first example (a) all ground-state
configurations are related by $1$-bit differences, while in the second
example (b) the Hamming distances between certain ground-state
configurations can be large---which  means that it takes longer for the
system to move from one ground-state configuration to another, therefore
causing larger fluctuations in the ground-state frequency.
Larger Hamming distances have been omitted for better visibility.
}
\label{fig:small_large}
\end{figure}

\begin{figure}[!htb]
\centering
\includegraphics[width=\columnwidth]{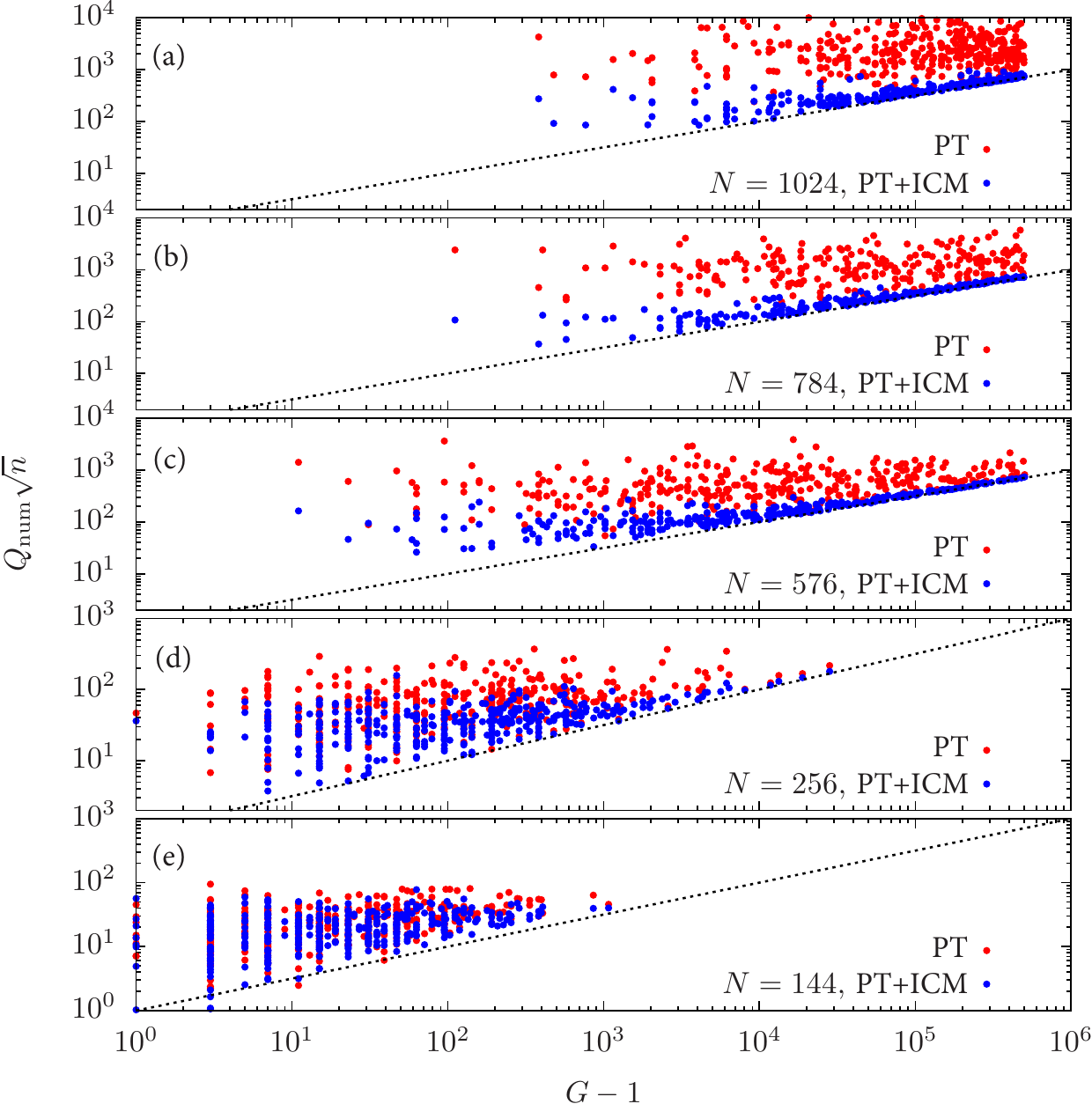}
\caption{
Scatter plot of $Q_{\rm num}\sqrt n$ as a function of the ground-state
degeneracy $G-1$ for different spin-glass instances with different
system sizes $N$ on a two-dimensional lattice. The data points for PT+ICM
(blue/dark color) are closer to the theoretical limit than those for 
PT (red/light color), and this improvement gets better as the system
size increases.
Data for (a) $N = 1024$,
(b) $N = 784$,
(c) $N = 576$,
(d) $N = 256$, and
(e) $N = 144$.
}
\label{fig:2d_scatter}
\end{figure}

\begin{figure}[!htb]
\centering
\includegraphics[width=\columnwidth]{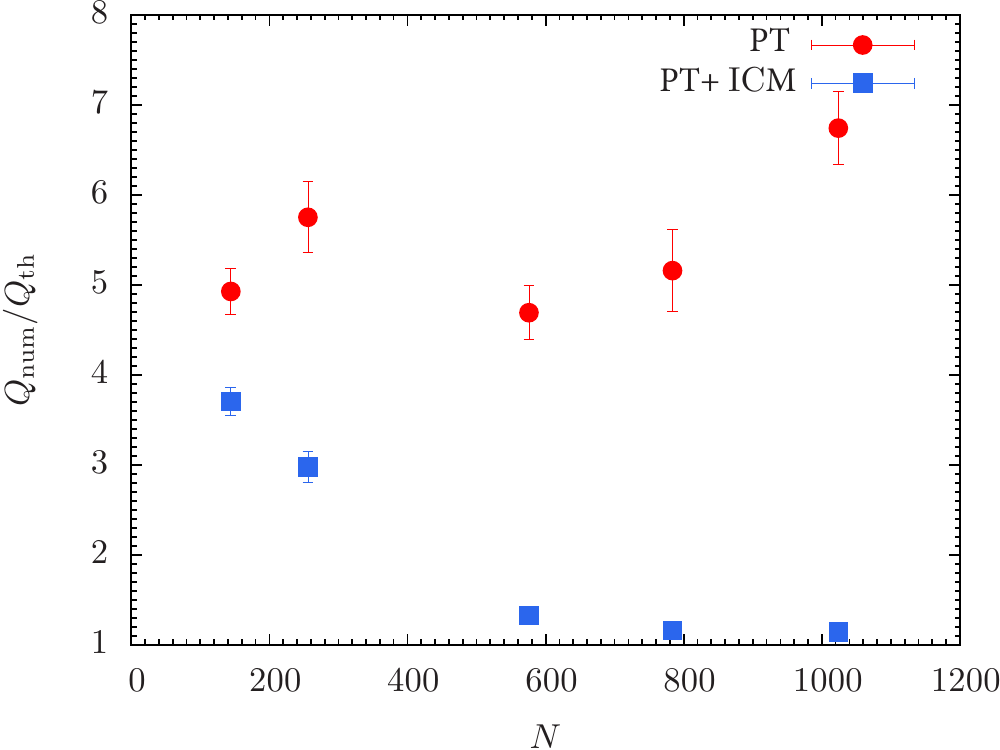}
\caption{
Median ratio $Q_{\rm num}/Q_{\rm th}$ for spin-glass instances as a
function of the system size $N$ on a two-dimensional lattice. The data
points show that PT+ICM (blue squares) performs better than PT (red
circles) for all system sizes and the gain is more significant with
increasing system size. Statistical error bars are
determined by a bootstrap analysis.
}
\label{fig:2d_ratio}
\end{figure}

In addition to studying how fair sampling behaves with increasing system
size, we also investigate how the quality of fair sampling is related to
ground-state degeneracy and plot $Q_{\rm num}/Q_{\rm th}$ as a function
of ground-state degeneracy for $N=800$ variables.  Figure
\ref{fig:chimera_Q_degeneracy} suggests that the more ground-state
configurations, the easier to sample all ground-state configurations
with near-equal probabilities. This is not surprising because a large
ground-state manifold makes the algorithm easier to explore the
configuration space. We do emphasize, however, that in cases where the
ground-state degeneracy is exponentially large and with limited
resources only a subset of minimizing configurations is accessible, and for
these the sampling improves, the more configurations are present.
Furthermore, careful examination of instances with the same system size
and ground-state degeneracy suggests that the $Q_{\rm num}/Q_{\rm th}$
ratio is closely related to the Hamming distances between ground-state
configurations. It is shown in Fig.~\ref{fig:chimera_Q_hamming} that the
Sidon instances \cite{katzgraber:15,zhu:16} where $J_{ij} \in
\{\pm5,\pm6,\pm7\}$ with large Hamming distances between the
ground-state configurations tend to have a high $Q_{\rm num}/Q_{\rm th}$
ratio compared to those with small Hamming distances between the states.
Here, PT+ICM achieves more equiprobable sampling with large Hamming
distances. Figure \ref{fig:small_large} shows two examples of
ground-state configurations with different Hamming distances on a
Chimera graph with $N=128$. PT+ICM's cluster updates allow nonlocal
moves in the energy landscape, therefore reducing $Q_{\rm num}/Q_{\rm
th}$ for instances with large Hamming distances between the ground-state
configurations.

Figure \ref{fig:2d_scatter} shows $Q_{\rm num}\sqrt n$ and $Q_{\rm
th}\sqrt n$ as a function of the ground-state degeneracy $G-1$ for
different spin-glass instances on a two-dimensional square lattice.
Similar to the Chimera graph case, the data using PT+ICM (blue/dark
color) are closer to the theoretical optimality line than the data using
PT (red/light color), and the discrepancy between PT+ICM and PT becomes
larger  as the system size increases. In Fig.~\ref{fig:2d_ratio}, the
median ratio $Q_{\rm num}/Q_{\rm th}$ again demonstrates that PT+ICM
is superior to PT in this case for square lattices.

\subsection{Estimating the ground-state degeneracy}

We also develop an approximate method to count the number of
ground-state configurations based on the fair sampling capabilities of
PT+ICM and compare the results to exact methods \cite{galluccio:00}
for a handful of configurations. Counting problems \cite{gomes:08}
typically ask how many solutions exist for a given instance and belong
to complexity class of \#P. This approximate method exploits the fact
that if one can sample ground states uniformly then one can obtain a
reasonable order-of-magnitude estimate of the ground-state degeneracy.
Our renormalization-inspired approach works as follows:
\begin{enumerate}

\item{Compute the ground-state energy $E_0$ for a fixed number of Monte
Carlo sweeps (see above).}

\item{Sample the number of ground states $G_0$ for the full system for a
fixed number of Monte Carlo sweeps.}

\item{Iteratively restrict the number of free variables (i.e.,
those that are not restricted) and estimate the ratio
$$R_{i-1}=G_{i-1}/G_{i}$$ for a fixed number of Monte Carlo sweeps.}

\item{Repeat until the system size is small enough to be able to compute
the number of ground state configurations $G_{\rm final}$ exactly, e.g., 
via enumeration.}

\item{Multiply the product of ratios by the exact count of
ground-state configurations to estimate the number of ground states for
the full system via $$G^{\rm RG} =G_{\rm
final}\prod\limits_{i}{R_{i-1}}.$$}

\end{enumerate}
We compare results of this approximate method to exact counts on a
two-dimensional square lattice with bimodal coupling constants $J_{ij}
\in \{\pm1\}$.  Simulation parameters and results are shown in Table
\ref{tab:gs_counting}. The renormalization-based estimates agree with
the exact ground-state degeneracy within error bars.

\begin{table}
\caption{
For each instance  with system size
$N=1024$, we run $N_{\rm sw} = 2^{23}$ Monte Carlo sweeps for each of
the $4 N_T=4 N_{hc}=120$ replicas with lowest temperature $T_{\rm
min}=0.17$ and highest temperature $T_{\rm max}=1.3$. The fixed number
of Monte Carlo sweeps and free variables for each iteration are $1/24
N_{\rm sw}$ and $1/24 N$, respectively. The median estimate of the
degeneracy $G^{\rm RG}$ is averaged over $10$ independent runs and error bars are
computed using the jackknife method.
\label{tab:gs_counting}
}
\begin{tabular*}{\columnwidth}{@{\extracolsep{\fill}} l c c c c }
\hline
\hline\\[-3mm]
Instance & $G^{\rm exact}$ & $G^{\rm RG}$& error & \% error\\
\hline
$0$ & $2.0094\times10^{29}$ & $1.9415\times10^{29}$  &  $\pm 6.04\times10^{27}$& $3.00\%$\\
$1$ & $9.7771\times10^{34}$ & $1.0081\times10^{35}$  &  $\pm 3.58\times10^{33}$& $3.66\%$\\ 
$2$ & $3.3778\times10^{27}$ & $3.3188\times10^{27}$  &  $\pm 1.18\times10^{26}$& $3.50\%$\\
$3$ & $1.2826\times10^{32}$ & $1.3041\times10^{32}$  &  $\pm 2.57\times10^{30}$& $2.00\%$\\
$4$ & $1.8613\times10^{39}$ & $1.9317\times10^{39}$  &  $\pm 7.59\times10^{37}$& $4.08\%$\\
$5$ & $1.4104\times10^{40}$ & $1.4515\times10^{40}$  &  $\pm 7.28\times10^{38}$& $5.16\%$\\
$6$ & $9.6510\times10^{29}$ & $9.6105\times10^{29}$  &  $\pm 1.83\times10^{28}$& $1.90\%$\\
$7$ & $2.3699\times10^{38}$ & $2.3543\times10^{38}$  &  $\pm 1.66\times10^{37}$& $7.04\%$\\
$8$ & $1.4168\times10^{31}$ & $1.3527\times10^{31}$  &  $\pm 1.16\times10^{30}$& $8.58\%$\\
$9$ & $1.1265\times10^{34}$ & $1.0789\times10^{34}$  &  $\pm 5.06\times10^{32}$& $4.69\%$\\
\hline
\end{tabular*}
\end{table}

\section{Conclusions}

We have demonstrated that PT+ICM---parallel tempering Monte Carlo with
isoenergetic cluster moves---samples ground-state configurations fairly
and is an ideal method for applications where a pool of diverse
solutions is needed.  We also find that degeneracy and Hamming distances
between different ground-state configurations are closely related to the
relative standard deviation of frequency with which the ground states
are found, namely: ground states with large degeneracy and small Hamming
distances have a lower relative standard deviation of frequency.  It
will be interesting to exploit near-uniform sampling for model counting
\cite{wei:05} and SAT filter construction \cite{weaver:14,douglass:15}
in the future.

\begin{acknowledgments} 

We would like to thank Ruben S.~Andrist, Hamid Khoshbakht, Martin Weigel
and Salvatore Mandr{\`a} for fruitful discussions. We especially thank
Martin Weigel for providing the exact number of ground-state
configurations on two-dimensional square lattices using the Vondrak
code, and Ruben S.~Andrist for rendering Fig.~5. H.G.K.~acknowledges
support from the National Science Foundation (Grant No.~DMR-1151387) and
would like to thank Banh Mi for providing the necessary motivation for
this research. We thank the Texas Advanced Computing Center (TACC) at
The University of Texas at Austin and Texas A\&M University for
providing HPC resources.  Part of this research is based upon work
supported in part by the Office of the Director of National Intelligence
(ODNI), Intelligence Advanced Research Projects Activity (IARPA), via
MIT Lincoln Laboratory Air Force Contract No.~FA8721-05-C-0002.  The
views and conclusions contained herein are those of the authors and
should not be interpreted as necessarily representing the official
policies or endorsements, either expressed or implied, of ODNI, IARPA,
or the U.S.~Government.  The U.S.~Government is authorized to reproduce
and distribute reprints for Governmental purposes notwithstanding any
copyright annotation thereon.

\end{acknowledgments}

\bibliography{refs,comments}

\begin{thebibliography}{58}
\bibitem[{com({\natexlab{a}})}]{comment:pa}
\bibinfo{note}{Note that population annealing and parallel tempering Monte
  Carlo are comparably efficient \cite{wang:15} solvers.}
\bibitem[{com({\natexlab{b}})}]{comment:ratio_one}
\bibinfo{note}{Note that for most instances the ratio $Q_{\rm num}/Q_{\rm th}$
  is indeed greater than one, However, there are a few exceptions with small
  system size and ground-state degeneracy where $Q_{\rm num}/Q_{\rm th}$ is
  less than one. These disparities are due to the limited CPU resources,
  because in theory all ground-state configurations are supposed to be found
  with the same probabilities at long-time limit.}
\end{thebibliography}

\end{document}